\documentclass[%
 reprint,
 superscriptaddress,
nofootinbib,
 amsmath,amssymb,
 aps,
 prd,
]{revtex4-1}

\usepackage{graphicx}
\usepackage{dcolumn}
\usepackage{bm}
\usepackage{hyperref}
\usepackage[mathlines]{lineno}

\usepackage{xcolor}

\begin{document}

\preprint{APS/123-QED}

\title{Effect on Dark Matter Exclusion Limits from New Silicon Photoelectric Absorption Measurements}

\affiliation{Institut  f{\"u}r  Experimentalphysik,  Universit{\"a}t  Hamburg,  22761  Hamburg,  Germany}
\affiliation{Department of Physics, University of Toronto, Toronto, ON M5S 1A7, Canada}
\affiliation{Department of Physics, Stanford University, Stanford, CA 94305 USA}
\affiliation{Department of Physics, Southern Methodist University, Dallas, TX 75275, USA}
\affiliation{Fermi National Accelerator Laboratory, Center for Particle Astrophysics, Batavia, IL 60510 USA} 
\affiliation{D\'epartement de Physique, Universit\'e de Montr\'eal, Montr\'eal, Qu\'ebec H3C 3J7, Canada}
\affiliation{Department of Physics, Harvard University, Cambridge, MA 02138 USA}
\affiliation{Kavli Institute for Particle Astrophysics and Cosmology, Menlo Park, CA 94025 USA}
\affiliation{Kavli Institute for Cosmological Physics, University of Chicago, Chicago, IL 60637, USA}

\author{B.~von~Krosigk}\email{belina.von.krosigk@uni-hamburg.de} \affiliation{Institut  f{\"u}r  Experimentalphysik,  Universit{\"a}t  Hamburg,  22761  Hamburg,  Germany}
\author{M.J.~Wilson}\email{mwilson@physics.utoronto.ca} \email{matthew.james.wilson@uni-hamburg.de} \affiliation{Department of Physics, University of Toronto, Toronto, ON M5S 1A7, Canada} \affiliation{Institut  f{\"u}r  Experimentalphysik,  Universit{\"a}t  Hamburg,  22761  Hamburg,  Germany}
\author{C.~Stanford}\affiliation{Department of Physics, Stanford University, Stanford, CA 94305 USA}\affiliation{Department of Physics, Harvard University, Cambridge, MA 02138 USA}
\author{B.~Cabrera} \affiliation{Department of Physics, Stanford University, Stanford, CA 94305 USA} \affiliation{Kavli Institute for Particle Astrophysics and Cosmology, Menlo Park, CA 94025 USA}
\author{R.~Calkins} \affiliation{Department of Physics, Southern Methodist University, Dallas, TX 75275, USA}
\author{D.~Jardin} \affiliation{Department of Physics, Southern Methodist University, Dallas, TX 75275, USA}
\author{N.A.~Kurinsky} \affiliation{Fermi National Accelerator Laboratory, Center for Particle Astrophysics, Batavia, IL 60510 USA} \affiliation{Kavli Institute for Cosmological Physics, University of Chicago, Chicago, IL 60637, USA}
\author{F.~Ponce} \affiliation{Department of Physics, Stanford University, Stanford, CA 94305 USA}
\author{C.-P.~Wu}\affiliation{D\'epartement de Physique, Universit\'e de Montr\'eal, Montr\'eal, Qu\'ebec H3C 3J7, Canada}

\date{\today}

\begin{abstract}
Recent breakthroughs in cryogenic silicon detector technology allow for the observation of single electron-hole pairs released via particle interactions within the target material. This implies sensitivity to energy depositions as low as the smallest band gap, which is $\sim1.2$\,eV for silicon, and therefore sensitivity to eV/$c^2$-scale bosonic dark matter and to thermal dark matter at masses below 100\,MeV/$c^2$. Various interaction channels that can probe the lowest currently accessible masses in direct searches are related to standard photoelectric absorption. In any of these respective dark matter signal models any uncertainty on the photoelectric absorption cross section is propagated into the resulting exclusion limit or into the significance of a potential observation. Using first-time precision measurements of the photoelectric absorption cross section in silicon recently performed at Stanford~University, this article examines the importance having accurate knowledge of this parameter at low energies and cryogenic temperatures for these dark matter searches.

\end{abstract}

\maketitle


\section{\label{sec:intro}Introduction}

A diverse set of astrophysical observations provides compelling evidence for the existence of dark matter (DM) \cite{Zyla:2020zbs} that accounts for about 85\% of the matter content in the universe \cite{planck18}. These data do not, however, give much insight into the particle nature of dark matter and its non-gravitational interactions. Lacking knowledge about defining properties of the dark matter particles, it is of great scientific interest to search for as many plausible dark matter candidates as possible in a diverse set of interaction channels. Widely accepted candidates include, but are not limited to, WIMPs (weakly interacting massive particles) \cite{Jungman_1996}, LDM (light dark matter) \cite{B_hm_2004}, ALPs (axion-like particles) \cite{Peccei:2006as,Svrcek:2006yi} and dark photons \cite{Holdom:1985ag}, candidates to which current and future direct detection experiments are particularly sensitive \cite{Battaglieri:2017aum, Schumann_2019}.

Several of these direct detection channels are similar in that they can be related to the Standard Model photoelectric absorption cross section $\sigma_{\mathrm{p.e.}}$. Relevant processes include the absorption of ALPs and dark photons by a target material \cite{Bloch:2016sjj, Hochberg:2016sqx} (causing the emission of one or more electrons), and inelastic nuclear scattering of WIMPs or LDM (resulting in Bremsstrahlung emission of a photon \cite{Kouvaris:2016afs} or the emission of an electron via the Migdal Effect \cite{Ibe:2017yqa}). The natural low-energy limit for these searches in solid state detector experiments is the lowest band gap or ionization energy of the target material; $\sigma_{\mathrm{p.e.}}$ approaches zero below this energy, where only negligible contributions to the cross section due to material impurities and free carrier absorption exist.

Recent breakthroughs in cryogenic silicon (Si) detector technology \cite{Romani:2017iwi, Tiffenberg:2017aac, Aguilar_Arevalo_2017, Hong_2020} allow for the potential observation of these absorption and scattering processes at energies as low as the indirect band gap at $\sim 1.2$\,eV \cite{Ramanathan_2020}, thus maximizing the reach of the corresponding dark matter searches. Precise and accurate knowledge of $\sigma_{\mathrm{p.e.}}$ near the Si band gap and at cryogenic temperatures thus comes to the fore. Lacking respective absorption data to date, the uncertainty in $\sigma_{\mathrm{p.e.}}$ has been the dominant source of uncertainty at low ALP and dark photon masses in recent Si dark matter search results \cite{Agnese_2018, Amaral:2020ryn}; this fact has been directly addressed by a new, first-time measurement of $\sigma_{\mathrm{p.e.}}$ with a cryogenic Si device operated well below 1K by \citet{pexsec_stanford}.

This article discusses the sensitivity of dark matter direct detection interaction channels to uncertainties in $\sigma_{\mathrm{p.e.}}$. We show that for some interaction channels the systematic uncertainties cannot be ignored, whereas in other channels the uncertainties are important only in special cases. The current status of low energy $\sigma_{\mathrm{p.e.}}$ measurements is summarized in Sec.~\ref{sec:pe}. The dark matter signal models associated with $\sigma_{\mathrm{p.e.}}$ are described in Sec.~\ref{sec:DMmodels}. The effect of the new $\sigma_{\mathrm{p.e.}}$ measurements on corresponding dark matter coupling limits is discussed in Sec.~\ref{sec:results}, followed by a conclusion in Sec.~\ref{sec:conclusion}.

\section{\label{sec:pe}Photoelectric absorption}
The direct detection of dark matter requires knowledge of the photoelectric absorption cross section $\sigma_{\mathrm{p.e.}}$ over a wide range of energies $E$, depending on the interaction channel and on the dynamic range of the experiment. For the discussed interaction channels and in leading-edge, low-threshold silicon detectors, energies as low as $\sim 1.2$\,eV and higher than 1\,keV are required. The nominal $\sigma_{\mathrm{p.e.}}(E)$ curve is defined as the $\sigma_{\mathrm{p.e.}}$ data from Ref.~\cite{pexseclit4} for energies $\leq 1$\,keV, Ref.~\cite{pexseclit0} for energies $> 1$\,keV and $\leq 20$\,keV, and Ref.~\cite{XCOM} for energies $> 20$\,keV. This definition roughly follows that used in Ref.~\cite{Hochberg:2016sqx}.

\begin{figure}[tbh!] 
    \centering
    \includegraphics[width=1.0\columnwidth]{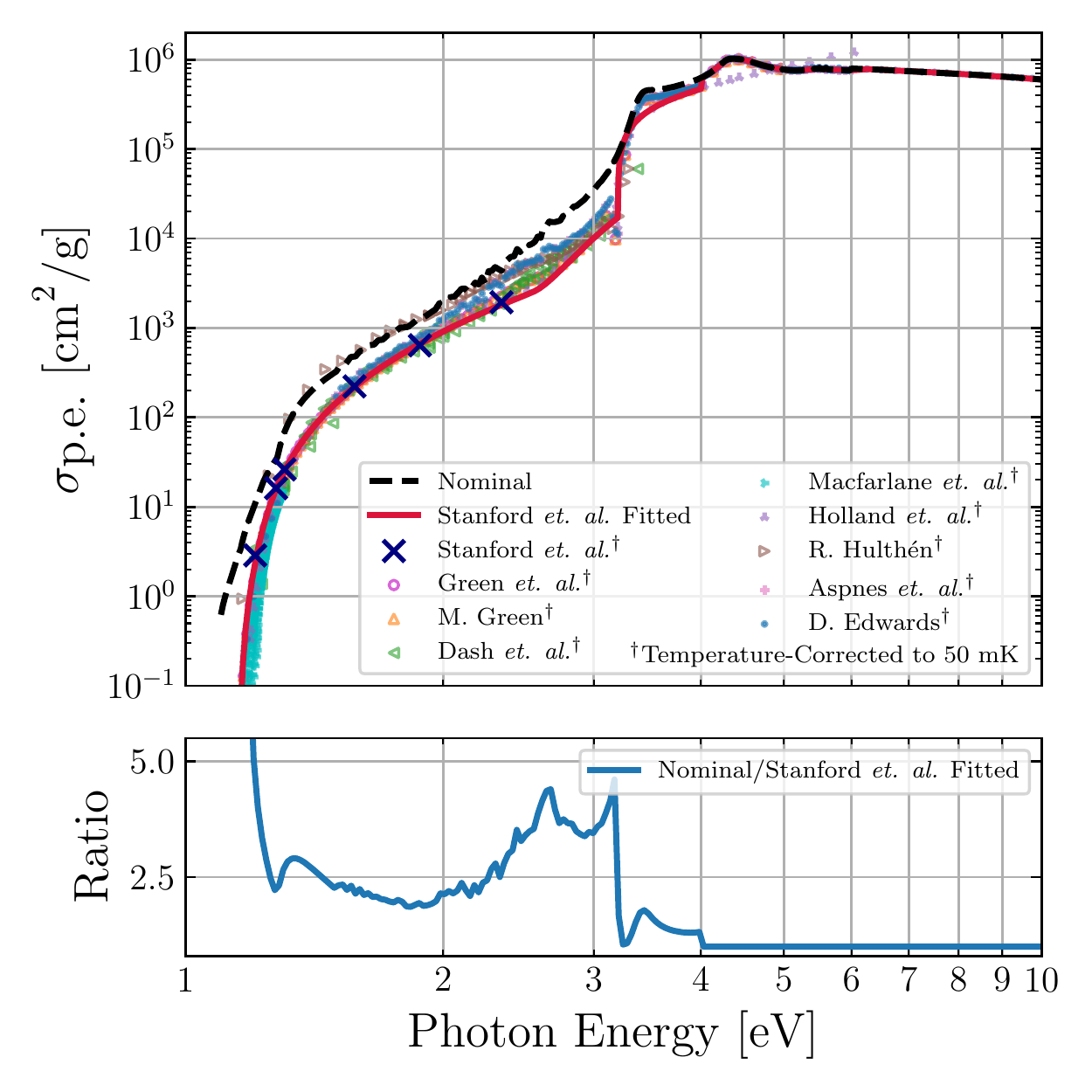}\\
    \caption{{\bf Top:} Summary of existing measurements of the photoelectric absorption cross section data at low energies \cite{pexseclit1,pexseclit2,pexseclit3,pexseclit4,pexseclit5,pexseclit6,pexseclit7,pexseclit8}. The data have been temperature-corrected to 50\,mK, using the model from Ref.~\cite{Rajkanan:1979}, with fitted parameters from Ref.~\cite{pexsec_stanford}. Also shown are the new measurements by \citet{pexsec_stanford} taken at 0.5\,K, also temperature-corrected to 50\,mK using the same method. Lastly, this plot shows the difference between the nominal and fitted $\sigma_{\mathrm{p.e.}}$ curves used to compare the limits in this work. The nominal curve is defined as described in the text. The fitted curve uses the model in Ref.~\cite{Rajkanan:1979} evaluated at 50\,mK with fit parameters extracted from the new data described in Ref.~\cite{pexsec_stanford}. The model is valid up to 4\,eV. Above that energy, the fitted curve and the nominal curve are the same. {\bf Bottom:} Ratio of the nominal $\sigma_{\mathrm{p.e.}}$ curve over the fitted $\sigma_{\mathrm{p.e.}}$ curve.}
    \label{fig:pexsec_showcase}
\end{figure}

At incident photon energies below $\sim$4\,eV, there is temperature dependence in the photoelectric cross section due to the temperature-dependent phonon distributions that are required for indirect, phonon-assisted photon absorption \cite{Rajkanan:1979}. Until recently, existing literature about $\sigma_{\mathrm{p.e.}}$ measurements at energies of $\mathcal{O}$(eV) \cite{pexseclit0,pexseclit1,pexseclit2,pexseclit3,pexseclit4,pexseclit5,pexseclit6,pexseclit7,pexseclit8} did not include measurements at temperatures below 5\,K, i.e. at temperatures highly relevant to state-of-the-art Si direct detection dark matter searches \cite{Agnese_2018, Amaral:2020ryn, SENSEI2019, DAMIC2019}. To account for this lack of data, higher temperature results had to be extrapolated down to the lower temperature region of interest. A notable systematic uncertainty of up to an order of magnitude on the cross section remained due to the wide spread in the existing data even after temperature corrections, as can be seen in Fig.~\ref{fig:pexsec_showcase}. It should be noted that in some of the sources shown the experimental setup differs, which could lead to systematic shifts in the measurement of $\sigma_{\mathrm{p.e.}}$ (for example, Ref.~\cite{pexseclit8} used epitaxial Si, and Ref.~\cite{pexseclit7} used n-type Si). 
Overall, the uncertainty in $\sigma_{\mathrm{p.e.}}$ yielded dominating uncertainties in some dark matter results.

These uncertainties motivated a dedicated measurement of $\sigma_{\mathrm{p.e.}}$ in Si for photon energies of 1.2—2.8\,eV and temperatures as low as 0.5\,K \cite{pexsec_stanford}. Based on this new data, Ref.~\cite{pexsec_stanford} models the Si absorption cross section up to 4\,eV using an analytic description validated up to that energy \cite{Rajkanan:1979}. 

To fully probe the dark matter models described in Sec.~\ref{sec:DMmodels},
the photoelectric absorption cross section is needed over a wide range of energies. Thus, for energies larger than 4\,eV, previously existing data is used starting at 4.02\,eV, interpolating in the region in between. The corresponding cross section above 4.02\,eV follows the same curve as nominally used in various Si direct dark matter search experiments and phenomenological studies \cite{Hochberg:2016sqx}. 

The resulting photoelectric cross section curve from 1.2\,eV to 50\,keV is provided as supplemental material \footnote{The files provided as supplemental material include pre-existing material and is added for convenience. The reader should make sure to include the original sources in the citations, i.e. \cite{pexseclit4, pexseclit0, XCOM, pexsec_stanford}, if the provided data file is used.} and is referred to as the {\it fitted} $\sigma_{\mathrm{p.e.}}$ curve throughout this paper. In contrast, the photoelectric cross section curve made up of commonly used data is the previously defined nominal $\sigma_{\mathrm{p.e.}}$ curve. Above 4\,eV both curves are identical, and therefore the effects on the DM interaction channels are a direct result of the temperature dependence of $\sigma_{\mathrm{p.e.}}$.

\section{\label{sec:DMmodels}Dark Matter signal models}
For many years, most direct-detection dark matter experiments have been optimized for WIMP searches using nuclear recoil events. However, the absence of a confirmed positive signal to date has increased the interest in other dark matter candidates and dark matter masses lower than the $\mathcal{O}$(GeV/$c^2$-TeV/$c^2$) window, i.e. beyond the standard WIMP paradigm. Recent detector developments using cryogenic Si detectors show they are particularly sensitive to low energy electron recoil events induced by dark matter candidates with masses as low as about 1\,eV/$c^2$. The respective interactions are described in this section. The observable rate in each case is a product of the cross section and the relic dark matter flux $\phi=\rho_\mathrm{DM} v / m_\mathrm{DM}$, where $\rho_\mathrm{DM}=0.3$\,GeV/cm$^3$ is the local dark matter density, and $v$ and $m_\mathrm{DM}$ are the dark matter velocity and mass respectively \cite{Weber_2010}.

\subsection{\label{ssec:absorption}Absorption}

If the energy of a bosonic relic dark matter candidate, like a sub-MeV/$c^2$ dark photon or ALP, exceeds the work function of a particular target, it may be absorbed by bound electrons in analogy to the photoelectric effect \cite{Pospelov:2008jk, Dimopoulos:1985tm}. In the case of cryogenic semiconductor detectors, the band gap plays the role of the work function, which is $\sim1.2$\,eV for Si at 0\,K. The associated excitation of an electron into the conduction band is detected as an electron recoil event with an energy equal to the total energy of the incoming particle.

For cold dark matter, this energy is a good approximation of the mass energy, which means that the dynamic range of the experiment equals the mass range accessible in these absorption processes. With a threshold as low as the band gap, existing cryogenic semiconductor experiments can thus probe parameter space down to a few eV/$c^2$ exceeding current astrophysical bounds with only moderate exposure \cite{Agnese_2018, Amaral:2020ryn, SENSEI2019, DAMIC2019}. 


\subsubsection{\label{sssec:alps}Axion-Like Particles}

The expected cross section $\sigma_a$ for the effective ALP--electron interaction can be related to 
$\sigma_{\mathrm{p.e.}}$ as per
\begin{equation}
\label{equ:alpabs}
    \sigma_a(E_a) = \sigma_{\mathrm{p.e.}}(E_a) \frac{g_{ae}^2}{\beta_a} \frac{3 E_a^2}{16 \pi \,\alpha\, m_e^2 c^4} \left( 1 - \frac{\beta_a^{2/3}}{3} \right),
\end{equation}
\noindent where $m_e$ is the mass of the electron, $E_a$ is the ALP's total energy, $\beta_a=v_a/c$ is its relativistic beta factor with velocity $v_a$ and speed of light $c$, $\alpha$ is the fine structure constant, and $g_{ae}$ is the axioelectric coupling of the ALP to the electrons \cite{Pospelov:2008jk, Fu:2017lfc}. For non-relativistic ALPs Eq.~\ref{equ:alpabs} reduces to
\begin{equation}
\label{equ:alpabs_relic}
    \sigma_a(m_a) = \sigma_{\mathrm{p.e.}}(m_ac^2) \frac{g_{ae}^2}{\beta_a} \frac{3 m_a^2}{16 \pi \,\alpha\, m_e^2},
\end{equation}
\noindent with $E_a = m_ac^2$ and $\beta_a \ll 1$. The consequent interaction rate in the detector of ALPs constituting all of the relic dark matter 
is
\begin{equation}
\label{equ:alps_rate}
    R_a(m_a) = \rho_\mathrm{DM} \sigma_{\mathrm{p.e.}}(m_ac^2) \frac{3 \,c}{16 \pi \,\alpha\, m_e^2} g_{ae}^2 \, m_a.
\end{equation}
Using this signal model and given a measured energy spectrum in the region of interest, a limit can be set on $g_{ae}$ as a function of ALP dark matter mass $m_{a}$.

\subsubsection{\label{sssec:dp}Dark Photons}

The kinetic mixing of dark photons $A^\prime$ to Standard Model photons enables an effective coupling to electrons, and with it the absorption of dark photons by atoms. The expected cross section for this process \cite{Bloch:2016sjj, Hochberg:2016sqx} is given by
\begin{equation}
\label{equ:dpabs}
    \sigma_{A^\prime}(E_{A^\prime}) =  \frac{\varepsilon_{\mathrm{eff}}^2}{v_{A^\prime}}\sigma_{\mathrm{p.e.}}(E_{A^\prime}) n \hbar c,
\end{equation}
\noindent where $E_{A^\prime}$ is the dark photon's total energy, $v_{A^\prime}$ is the dark photon's velocity, $\varepsilon_{\mathrm{eff}}$ is the effective kinetic mixing parameter, and $n$ is the index of refraction. For dark photon masses $\gtrsim 20$\,eV/$c^2$ $\varepsilon_{\mathrm{eff}}$ approximates the kinetic mixing parameter $\varepsilon$, the actual parameter of interest. 

At lower masses in-medium effects can significantly alter $\varepsilon$ and it has to be derived from $\varepsilon_{\mathrm{eff}}$ using
\begin{equation}
    \label{equ:eps}
    \varepsilon^{2}_{\mathrm{eff}}=\frac{\varepsilon^{2} m^{2}_{A'}}{(m^{2}_{A'} - 2m_{A'}\sigma_{2} + \sigma_{2}^{2} + \sigma_{1}^{2})},
\end{equation}
\noindent as described in Ref.~\cite{Hochberg:2016sqx}. Here $\sigma_1$ and $\sigma_2$ are the energy-dependent real and imaginary part of the complex conductivity, respectively. The photoelectric cross section is related to the real part of the complex conductivity through:
\begin{equation}
\label{equ:sig1}
\sigma_{1}(m_{A'}) = n\cdot \sigma_{\textrm{p.e.}}(m_{A'})\cdot \rho \cdot \hbar c ,
\end{equation}

\noindent where $\rho$ is the density of the target material. Assuming that all relic dark matter consist of non-relativistic dark photons, the event rate is given by:
\begin{equation}
    \label{equ:eps_rate}
    R_{A^\prime}(m_{A^\prime}) = \frac{\rho_\mathrm{DM}}{m_{A^\prime}} \varepsilon_{\mathrm{eff}}^2 \sigma_{\textrm{p.e.}}(m_{A^\prime} c^2) n \hbar c.
\end{equation}
Using this signal model and given a measured energy spectrum in the region of interest, a limit can be set on $\varepsilon$ as a function of dark photon dark matter mass $m_{A^\prime}$.

\subsection{\label{ssec:brems}Bremsstrahlung}

The scattering of a thermal, relic dark matter particle $\chi$, such as a WIMP or LDM, with a target nucleus $N$ has both elastic and inelastic contributions. The inelastic scattering process is accompanied by an emitted photon and referred to as Bremsstrahlung \cite{Kouvaris:2016afs}.

While the total cross section of the Bremsstrahlung process is orders of magnitude lower than that of the elastic process, it allows for the search of dark matter at masses well below those accessible in elastic scattering searches. 
If the dark matter mass is $m_\chi \ll m_N$, the kinetic energy of the dark matter that gets transferred to the emitted Bremsstrahlung photon is much higher than the elastic nuclear recoil energy would be. This results in an observable electron recoil signature in the detector down to a mass of $m_\chi < 90$\,MeV/$c^2$, the lowest mass probed to date in elastic DM-nucleus scattering searches \cite{alkhatib2020light}.

The differential cross section as a function of emitted photon energy $E_{\gamma}$ can be written as
\begin{equation}
\label{equ:brems_dsigdE}
    \small
	\frac{d\sigma}{dE_{\gamma}} = 
	\frac{4\alpha |f(E_{\gamma})|^2}{3\pi E_{\gamma}} \frac{\mu_N^2 v^2\sigma_N^{\mathrm{SI}} }{m_N^2} 
	\sqrt{1-\frac{2E_{\gamma}}{\mu_N v^2} } \left( 1 - \frac{E_{\gamma}}{\mu_N v^2}   \right)
\end{equation}
\noindent where $\alpha$ is the fine structure constant, $f$ is the complex atomic scattering function, $\mu_N$ is the DM-nucleus reduced mass, and $m_N$ is the mass of the nucleus \cite{Kouvaris:2016afs}. The spin-independent elastic DM-nucleus scattering cross section $\sigma_N^{\mathrm{SI}}$ is related to the respective DM-nucleon cross section $\sigma_n^{\mathrm{SI}}$ through 
\begin{equation}
  \sigma_N^{\mathrm{SI}} \simeq A^2 \sigma_n^{\mathrm{SI}} \left( \frac{\mu_n}{\mu_N} \right)^2
\end{equation}
\noindent where $\mu_n$ is the DM-nucleon reduced mass.

Following the convention in Ref.~\cite{XRayDataBooklet}, 
$f$ can be written as a function of the photoelectric cross section:
\begin{equation}
|f|^2 = |f_1 + if_2|^2 = f_1^2 + f_2^2
\end{equation} 
\noindent where the imaginary part $f_2$ is
\begin{equation}
f_2(E_\gamma) = \frac{\sigma_\mathrm{p.e.}(E_\gamma)}{2r_e \lambda}
\end{equation} 
\noindent and where the real part $f_1$ relates to the imaginary part and thus the photoelectric cross section as
\begin{align}
f_1(E_\gamma) &= Z^* + \frac{2}{\pi} \mathcal{P} \int_{0}^{\infty} \frac{E^\prime_\gamma f_2(E^\prime_\gamma)}{E_\gamma^2 - E^{\prime 2}_\gamma} dE^\prime_\gamma \\\nonumber
&= Z^* + \frac{1}{\pi r_e hc} \mathcal{P} \int_{0}^{\infty} \frac{E^{\prime 2}_\gamma \sigma_\mathrm{p.e.}(E^\prime_\gamma)}{E_\gamma^2 - E^{\prime 2}_\gamma} dE^\prime_\gamma.
\end{align} 
Here, $Z^* \simeq Z-(Z/82.5)^{2.37}$ is the atomic number after a small relativistic correction, $r_e$ is the electron radius, $h$ is Planck's constant, $\lambda$ is the wavelength, and $\mathcal{P}$ is the Cauchy principal value.

The energy spectrum of an event in the detector is obtained by multiplying the cross section by the number of target nuclei per unit mass $N_T$ and the relic dark matter flux, averaging the cross section over the dark matter velocity distribution $f_v$ in the lab frame:
\begin{equation}
\label{equ:brems_dRdE}
	\frac{dR}{dE_{\gamma}} = N_T \frac{\rho_\mathrm{DM}}{m_\chi} \int_{\left| \vec{v} \right| \geq v_{\mathrm{min}}} v f_v\left(\vec{v} + \vec{v}_e\right) \frac{d\sigma}{dE_{\gamma}} d^3\vec{v}.
\end{equation}
A truncated Maxwell-Boltzmann distribution with a most probable velocity of 220\,km/s is chosen for $f_v\left(\vec{v}\right)$. $\vec{v}_e$ is the velocity of the Earth relative to the galactic rest frame and $v_{\mathrm{min}} = \sqrt{2E_\gamma/\mu_N}$. 

The validity of this model at energies as low as the band gap is subject to ongoing research \cite{rouven}. Under this caveat, using this signal model and given a measured energy spectrum in the region of interest, a limit can be set on $\sigma^\mathrm{SI}_n$ as a function of dark matter mass $m_\chi$.



\subsection{\label{ssec:migdal}Migdal Effect}
Another process relevant to WIMP and LDM searches at masses too low for standard DM-nucleus elastic scattering experiments is the so-called Migdal Effect \cite{Migdal:1941}. In cases where the incoming kinetic energy is too small to fully dislocate a target nucleus from its host atom, the nucleus may at least temporarily be moved out of place. The surrounding electron cloud is not able to immediately follow the recoiling nucleus and the relative displacement within the atom represents an excited state. An electron emitted in the de-excitation process of the atom provides an observable electron recoil signal in the detector. The overall process is referred to as the Migdal Effect. Ref.~\cite{Ibe:2017yqa} describes how this process can be taken advantage of in direct dark matter detection experiments to increase the sensitivity to low-mass thermal dark matter. Various approaches exist for calculating the differential rate of this dark matter signal in different target materials \cite{Ibe:2017yqa,Liu:2020pat,Essig_2020,Baxter_2020,knapen2020migdal}. For the present paper only the approach in Ref.~\cite{Liu:2020pat} is relevant, as it is the only calculation based on $\sigma_{\mathrm{p.e.}}$.



in Ref.~\cite{Liu:2020pat}, the differential cross section for DM-nucleus scattering with nuclear recoil energy $E_R$ accompanied by an ionization electron of energy $E_r$ due to the Migdal Effect is related to the photoelectric cross section, as per
\begin{equation}
    \frac{d^2\sigma^{\mathrm{MPA}}}{dE_R dE_r} = \frac{m_e^2}{\mu_N^2v^2} \sigma^\mathrm{SI}_N\frac{E_R}{E_r}\frac{\sigma_{\mathrm{p.e.}}(E_r)}{4\pi^2\alpha}.
\end{equation}
This relation is referred to as "Migdal-photo-absorption" (MPA) relation.
The respective differential rate is obtained in the same fashion as in Sec.~\ref{ssec:brems} yielding
\begin{equation}
\label{equ:migdal_dRdEdE}
    \small
	\frac{d^2R}{dE_R dE_r} = N_T \frac{\rho_\mathrm{DM}}{m_\chi} \int_{\left| \vec{v} \right| \geq v_{\mathrm{min}}} v f_v\left(\vec{v} + \vec{v}_e\right) \frac{d\sigma^{\mathrm{MPA}}}{dE_R dE_r} d^3\vec{v}
\end{equation}

\noindent with $v_{\mathrm{min}} = (m_N E_R+\mu_N E_r)/(\mu_N\sqrt{2 m_N E_R}) $. For the case that the observed energy does not include the nuclear recoil energy $E_R$, the double differential rate can be reduced to
\begin{equation}
\label{equ:migdal_dRdE}
	\footnotesize
	\frac{dR}{dE_r} = N_T \frac{\rho_\mathrm{DM}}{m_\chi} \int dE_R \int_{\left| \vec{v} \right| \geq v_{\mathrm{min}}} d^3\vec{v} \,\, v f_v\left(\vec{v} + \vec{v}_e\right) \frac{d\sigma^{\mathrm{MPA}}}{dE_R dE_r}
\end{equation}
\noindent by integrating over $E_R$. Under the same caveat as in Sec.~\ref{ssec:brems}, using this signal model and a measured energy spectrum in the region of interest, a limit can be set on $\sigma^\mathrm{SI}_n$ as a function of dark matter mass $m_\chi$.

\section{\label{sec:results}Exclusion limits}
To demonstrate the effect of these new Si photoelectric absorption measurements on dark matter exclusion limits, we need to analyze low background data at electron recoil energies as low as the Si band gap. An excellent test case is provided by the SuperCDMS collaboration in their publication of the HVeV Run\,2 data analysis \cite{Amaral:2020ryn}. The same data and limit setting procedure is used here, summarized in this section. The HVeV Si device was operated at a temperature of about 50\,mK.

\subsection{Underlying Data and Toy Assumptions}
\label{ssec:data}

\begin{figure}[htb!] 
    \centering
    \includegraphics[width=1.0\columnwidth]{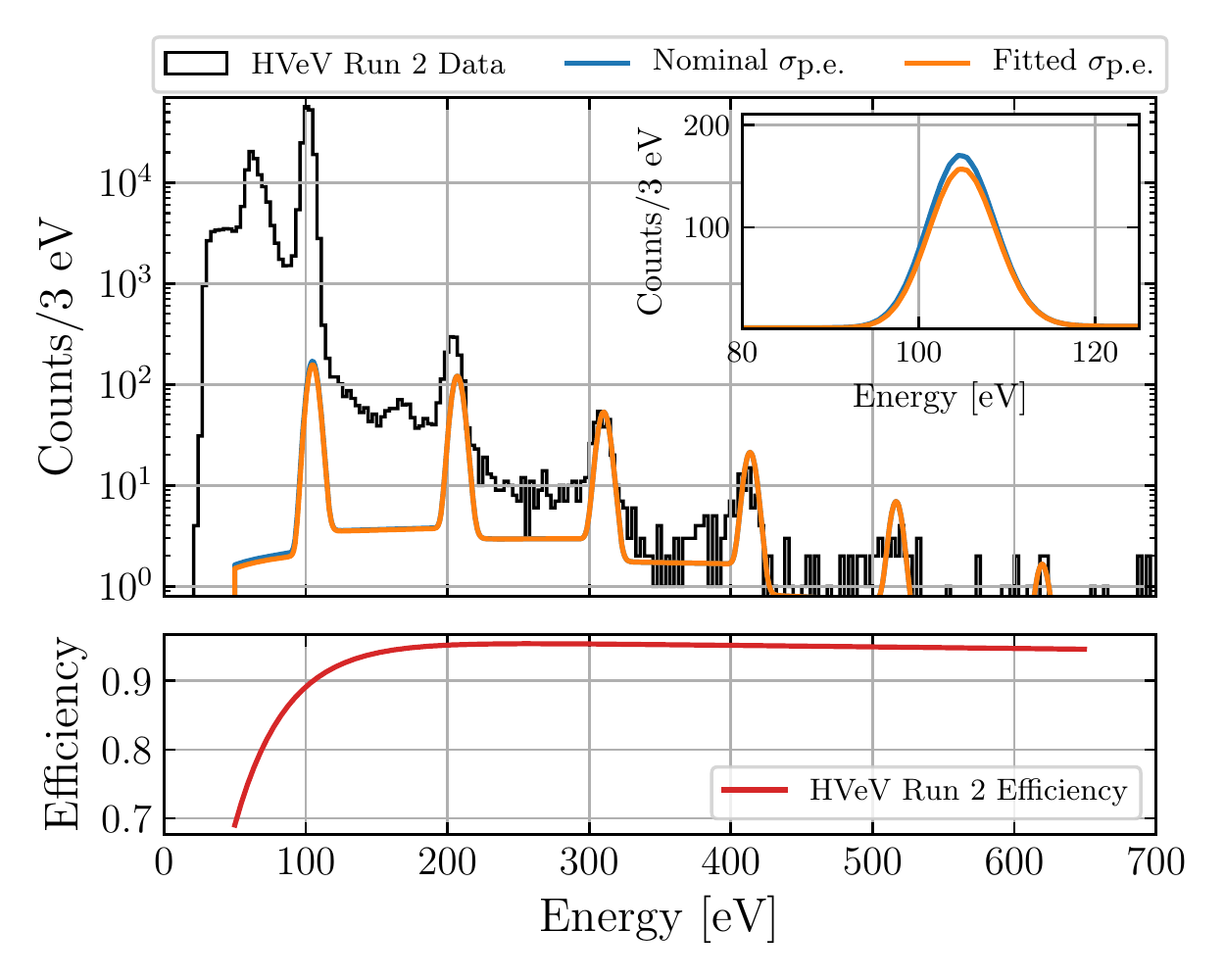}\\
    \caption{Data spectrum ({\bf top}, black histogram) and signal efficiency curve ({\bf bottom}) as a function of energy from the HVeV Run\,2 100\,V analysis \cite{Amaral:2020ryn}. The peaks just above 100\,V, 200\,V, 300\,eV, etc. are the quantization peaks at 1, 2, 3, etc. electron-hole pairs, respectively. The highest peak is at 6 electron-hole pairs. Also shown are the expected Migdal model spectra for an exemplary dark matter mass of 0.1\,GeV/$c^2$ and a cross section of $\sigma_n^\mathrm{SI}=1.36\times 10^{-30} $\,cm$^2$ for both $\sigma_\mathrm{p.e.}$ curves ({\bf top}, solid blue and orange lines). The inset shows the first electron-hole pair peak at which the difference between the two dark matter signal spectra is largest.}
    \label{fig:R2spec}
\end{figure}

The HVeV Run\,2 electron recoil spectrum and signal efficiency from Ref.~\cite{Amaral:2020ryn} are shown in Fig.~\ref{fig:R2spec}. The data were taken with an applied bias voltage of 100\,V. The exposure after analysis cuts is 1.219\,g$\cdot$day, the energy resolution is $\sigma_E=3.6$\,eV, the charge trapping fraction is 0.11, and the impact ionization fraction is 0.02. Throughout the presented analysis the fitted band gap energy of 1.13\,eV determined in Ref.~\cite{pexsec_stanford} is used, which is the only difference to the HVeV Run\,2 analysis in which 1.2\,eV was used as the band gap energy \cite{Amaral:2020ryn}. The experiment was operated within a surface facility at Northwestern University.

In the case of the Bremsstrahlung and Migdal model, an additional toy experiment is adduced as an example: in this idealized experiment, there are no events in the data spectrum, the efficiency is unity at all energies, and there is no trapping or impact ionization. The observable signature of the scattering signal models is less sensitive than the signature of the absorption signal models to differences in $\sigma_{\mathrm{p.e.}}$. This additional toy experiment maximizes the effect on the limits due the discussed differences in $\sigma_{\mathrm{p.e.}}$.

The detector response model used in this analysis is identical to the one used in Ref.~\cite{Amaral:2020ryn} to compute the mean number and probability distribution of electron-hole pairs produced in an interaction with a given deposited energy using a Fano factor value of 0.155. The detector response model also accounts for the effects of charge trapping and impact ionization, and was used to quantize the expected dark matter signal model at each mass. 

\subsection{Limit Calculation}
\label{ssec:poisson}

For this analysis, we compute limits using Poisson statistics under a signal-only hypothesis. For a given signal model $SM_0(E)$ calculated at a reference cross section $\sigma_0$, the cross section $\sigma_P$ (computed within an energy window $[a,b]$, with signal efficiency $\epsilon(E)$, and exposure $X$) is
\begin{equation}
\label{equ:poisson}
\sigma_{P} = \frac{C^{ab}_{P}\cdot \sigma_{0}}{X \int_{a}^{b} SM_{0}(E) \cdot \epsilon(E) dE},
\end{equation}
where $C^{ab}_{P}$ is the Poisson upper limit at a given confidence level (C.L.) on the event count between a and b. To account for the look-elsewhere effect, the C.L. used to calculate the limit at each window is adjusted such that the overall C.L. when selecting between $n$ number of windows is 90\%. For choosing between $n$ windows and an overall C.L. of 90\%, this correction is $0.9^{1/n}$. The windows used in this analysis are centered around the quantized signal peaks with a width of $\pm \,3\,\sigma_E$ in agreement with Ref.~\cite{Amaral:2020ryn}. 

 To compute the limits on $\varepsilon$ and $g_{ae}$, the same procedure is followed as in Ref.~\cite{Amaral:2020ryn}. For the $\sigma_n^{\mathrm{SI}}$ exclusion limits using HVeV Run\,2 data, all six peaks are taken into account. In case of the zero background toy experiment, only the first electron-hole pair peak is used to calculate the $\sigma_n^{\mathrm{SI}}$ limits, which is the peak with the greatest sensitivity to the differences in $\sigma_{\mathrm{p.e.}}$.

\subsection{Results}

\subsubsection{Dark Photon and ALP Absorption}
\label{sssec:abs_results}
\begin{figure*}[htb!] 
    \centering
    \includegraphics[width=1.0\columnwidth]{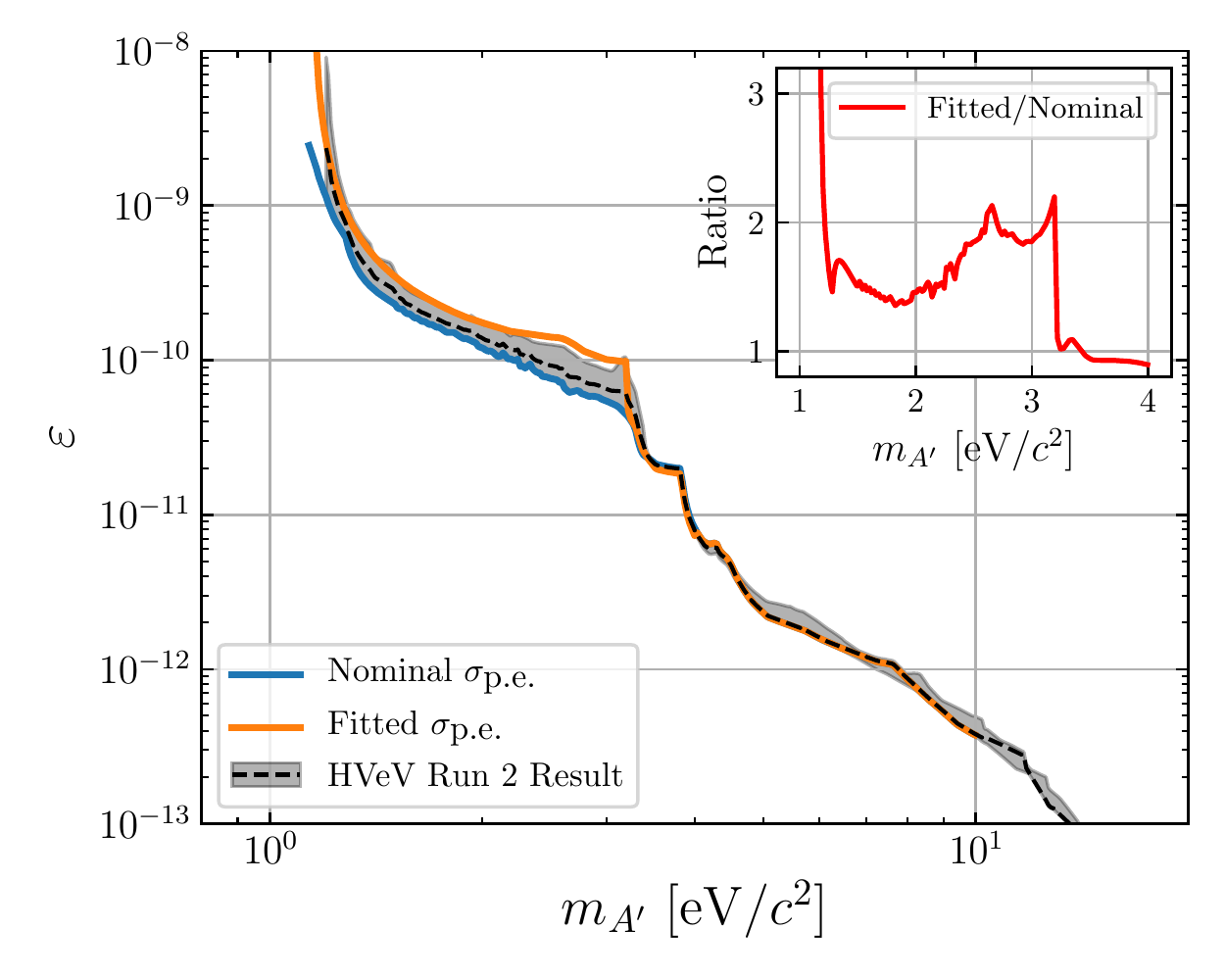}
    \includegraphics[width=1.0\columnwidth]{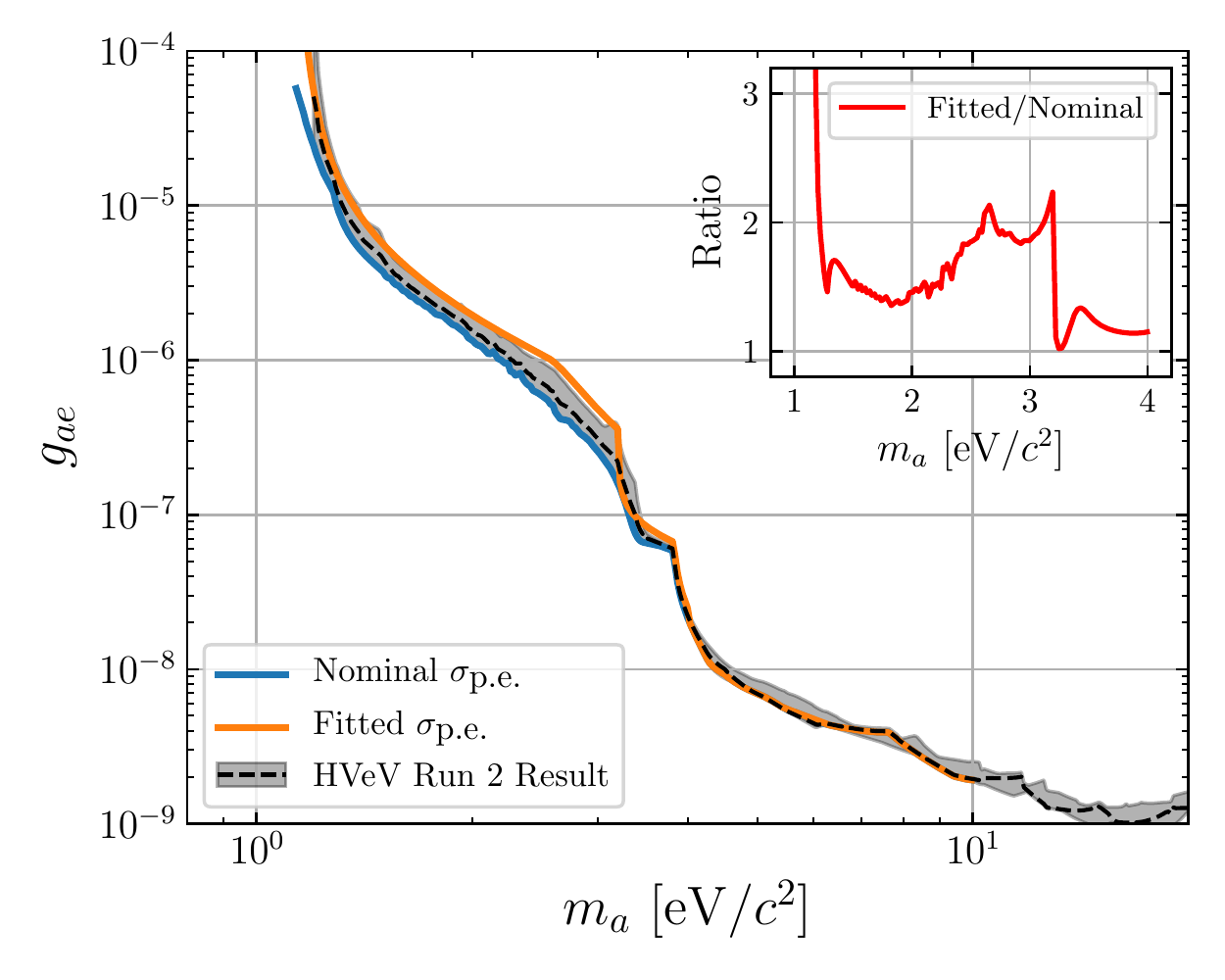}\\
    \caption{90\% C.L. exclusion limits on the dark photon kinetic mixing parameter $\varepsilon$ over the dark photon mass $m_{A'}$ ({\bf left}) and on the effective ALP-electron coupling constant $g_{ae}$ over the ALP mass $m_a$ ({\bf right}). All limits are based on SuperCDMS HVeV Run\,2 data \cite{Amaral:2020ryn}. The blue limit assumes the {\it nominal} dependence of $\sigma_{\mathrm{p.e.}}$ on energy. The orange limit takes into account the $\sigma_{\mathrm{p.e.}}$ curve {\it fitted} to recent direct measurements at sub-Kelvin temperatures in the energy range of interest \cite{pexsec_stanford} and evaluated at 50\,mK. Both limits, blue and orange, consider the band gap energy of $\sim 1.13$\,eV that was fitted in Ref.~\cite{pexsec_stanford}. Their ratio is shown in the inset on a linear scale. The limit in black is the published SuperCDMS HVeV Run\,2 result including its uncertainty band in gray.}
    \label{fig:limit_abs}
\end{figure*}
The signal generated in each of the bosonic dark matter absorption processes described in Sec.~\ref{ssec:absorption} is a delta function at the dark matter mass. In addition, the total event rate in both Eq.~\ref{equ:alps_rate} and Eq.~\ref{equ:eps_rate} depends on the photoelectric cross section at that mass. Therefore, a different $\sigma_{\mathrm{p.e.}}$ value at a given mass amounts to a difference in overall scaling of the expected dark matter signal spectrum for a given ALP coupling strength $g_{ae}$ or dark photon kinetic mixing $\varepsilon$. In both cases a lower $\sigma_{\mathrm{p.e.}}$ results in a higher, and thus weaker, limit on $g_{ae}$ or $\varepsilon$.

The notable difference between the nominal $\sigma_{\mathrm{p.e.}}$ curve and the newly fitted $\sigma_{\mathrm{p.e.}}$ curve, as shown in Fig.~\ref{fig:pexsec_showcase}, results in a significant shift in the exclusion limits. Figure~\ref{fig:limit_abs} compares the 90\% C.L. limits computed using either the nominal or fitted $\sigma_{\mathrm{p.e.}}$ curve. Shown for comparison are the published HVeV Run\,2 limits \cite{Amaral:2020ryn}. They differ below 4\,eV/$c^2$ from the nominal limits because the spread in the $\sigma_{\mathrm{p.e.}}$ data shown in Fig.~\ref{fig:pexsec_showcase} is propagated as systematic uncertainty in the limit. The HVeV Run\,2 limits are slightly weaker compared to the nominal limits because the HVeV Run\,2 limits incorporate
$\sigma_{\mathrm{p.e.}}$ information with lower values than the nominal curve.

The plot inset shows the ratio between the two newly calculated limits up to 4\,eV/c$^2$. The limit ratio uses the limit calculated with the nominal $\sigma_{\mathrm{p.e.}}$ in the denominator whereas the $\sigma_{\mathrm{p.e.}}$ ratio shown in Fig.~\ref{fig:pexsec_showcase} uses the nominal $\sigma_{\mathrm{p.e.}}$ in the numerator. A direct comparison of the ratio plots demonstrates their similarity and the overall scaling effect described above. The distinct features apparent in the limit ratio plots follow the behavior of the $\sigma_{\mathrm{p.e.}}$ ratio plot. Above a mass of 4\,eV/c$^2$ there is no difference between the limits as the same  $\sigma_{\mathrm{p.e.}}$ information is used.

The largest effect that the newly fitted $\sigma_{\mathrm{p.e.}}$ curve has on the resulting limits is closest to the band gap, where the ratio exhibits an asymptotic behavior. However even at masses larger than 1.2\,eV/$c^2$, throughout the entire region up to 4\,eV/$c^2$ in which the fitted $\sigma_{\mathrm{p.e.}}$ differs from the nominal $\sigma_{\mathrm{p.e.}}$, the shift is non-negligible. The limits are up to a factor of two weaker using the fitted compared to the nominal $\sigma_{\mathrm{p.e.}}$ curve. The only exception is the $\varepsilon$ limit at dark photon masses above about 3.5\,eV/$c^2$. In this region the limit based on the fitted $\sigma_{\mathrm{p.e.}}$ is slightly stronger due to the in-medium correction in Eq.~\ref{equ:eps} applied to $\varepsilon$ that also depends on $\sigma_{\mathrm{p.e.}}$. Overall the effect the recent $\sigma_{\mathrm{p.e.}}$ measurements have on the dark photon and ALP absorption searches with state-of-the-art cryogenic Si detectors cannot be ignored.

\subsubsection{Bremsstrahlung and Migdal Processes}
\label{sssec:scat_results}

\begin{figure*}[htb!] 
    \centering
    \includegraphics[width=1.0\columnwidth]{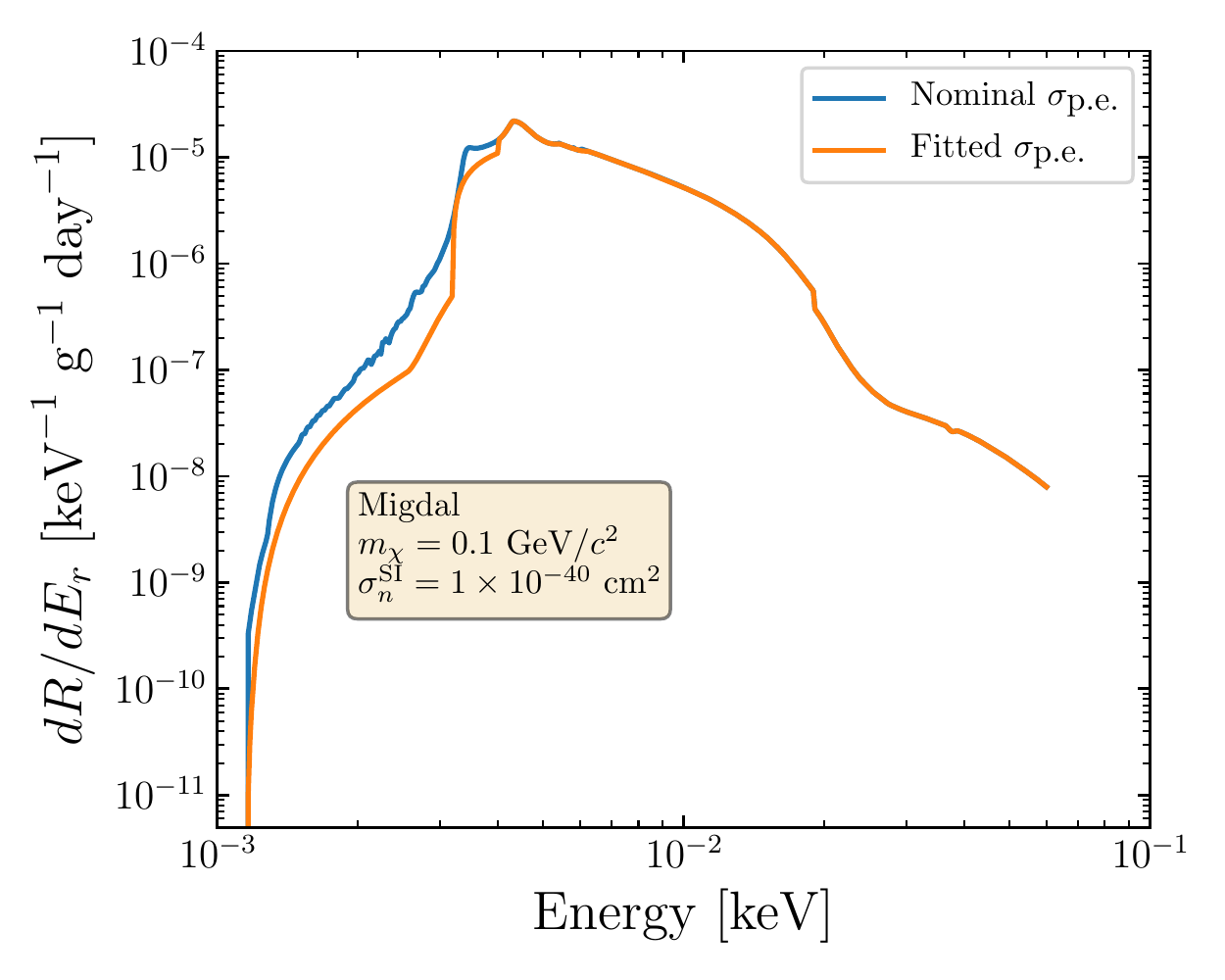}
    \includegraphics[width=1.0\columnwidth]{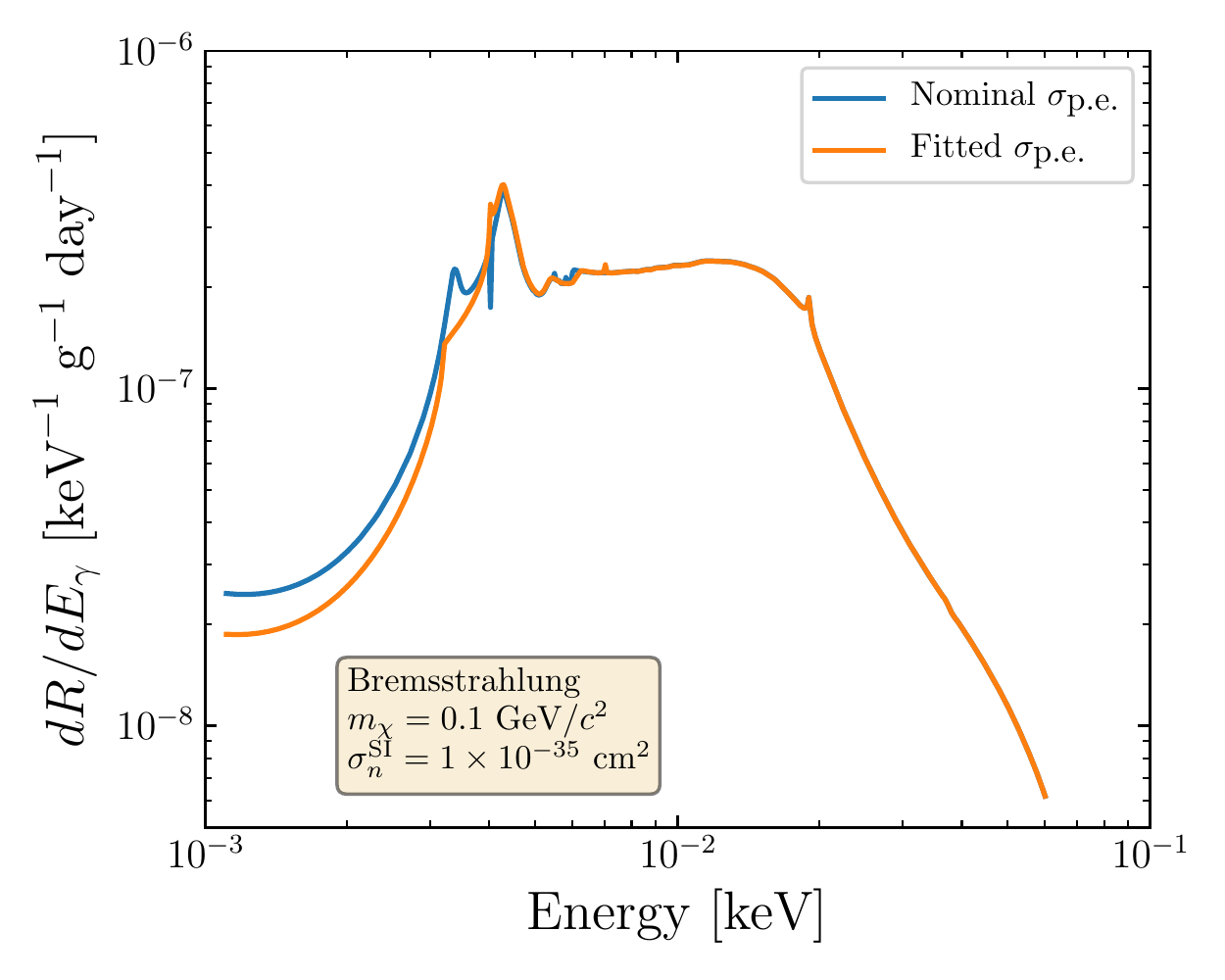}
    \caption{Differential interaction rate of inelastic DM-nucleus scattering under the emission of a Migdal electron ({\bf left}) or Bremsstrahlung photon ({\bf right}). The energy is the energy of the emitted particle. The assumed DM-nucleon scattering cross section $\sigma^\mathrm{SI}_n$ and dark matter mass $m_\chi$ are given in the text box. The differential rate is calculated once with the nominal $\sigma_{\mathrm{p.e.}}$ curve (blue) and once with the $\sigma_{\mathrm{p.e.}}$ curve fitted to the measurements in Ref.~\cite{pexsec_stanford} and evaluated at 50\,mK (orange).}
    \label{fig:sig_comp}
\end{figure*}

\begin{figure*}[htb!] 
    \centering
    \includegraphics[width=1.0\columnwidth]{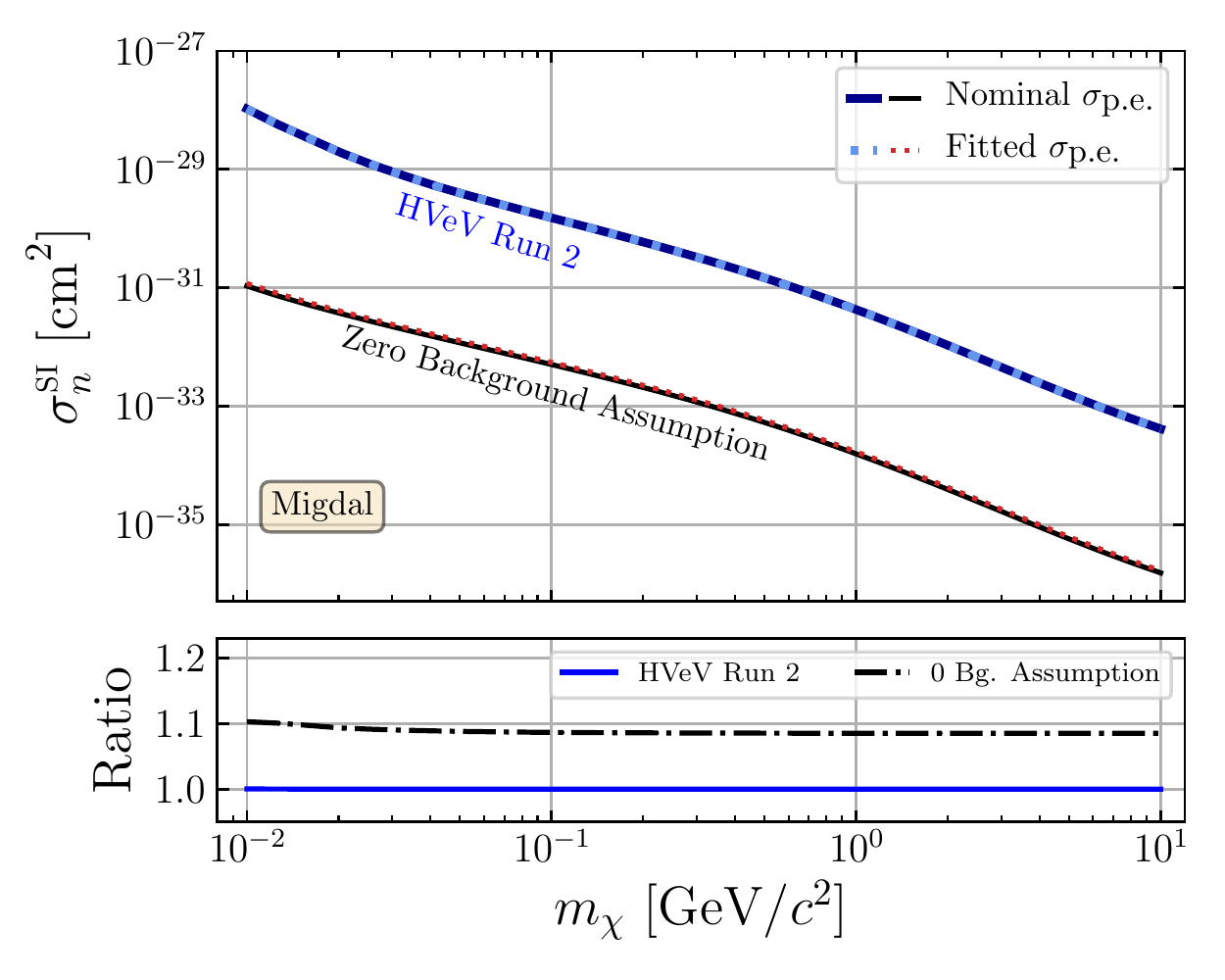}
    \includegraphics[width=1.0\columnwidth]{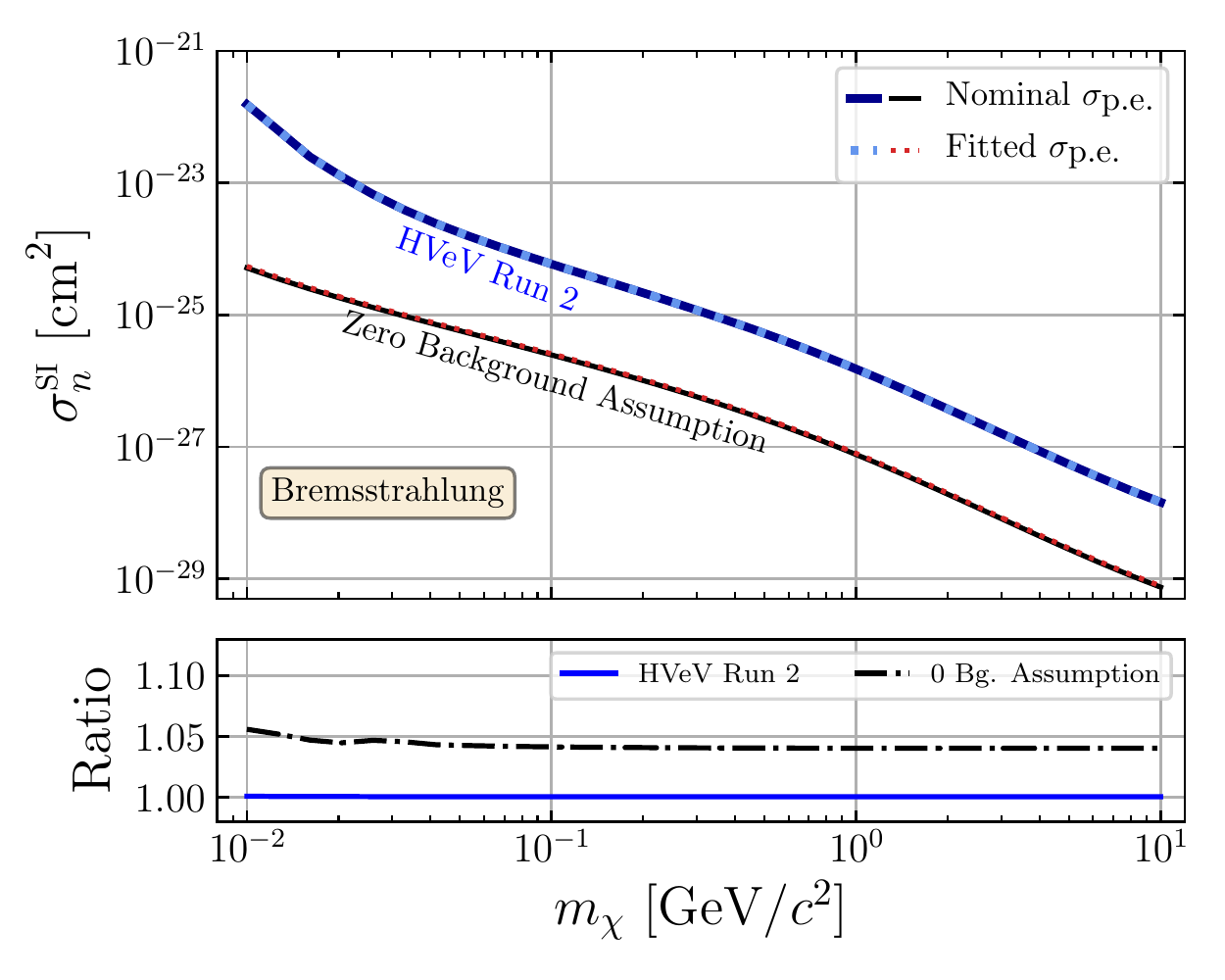}
    \caption{{\bf Top:} 90\% C.L. exclusion limits on the spin-independent DM-nucleon scattering cross section $\sigma^\mathrm{SI}_n$ over the dark matter mass $m_{\chi}$. The underlying interaction is inelastic DM-nucleus scattering under the emission of a Migdal electron ({\bf left}) or Bremsstrahlung photon ({\bf right}). The solid limits assume the {\it nominal} dependence of $\sigma_{\mathrm{p.e.}}$ on energy. The dotted limits take into account the $\sigma_{\mathrm{p.e.}}$ curve {\it fitted} to recent direct measurements at sub-Kelvin temperatures in the energy range of interest \cite{pexsec_stanford} and evaluated at 50\,mK. Two sets of limits are shown: one based on SuperCDMS HVeV Run\,2 data \cite{Amaral:2020ryn} and another based on a zero-background toy experiment. All limits consider the band gap energy of $\sim 1.13$\,eV that was fitted in Ref.~\cite{pexsec_stanford}. {\bf Bottom:} For both scenarios the ratio of the limit using the fitted $\sigma_{\mathrm{p.e.}}$ curve over the limit using the nominal $\sigma_{\mathrm{p.e.}}$ curve is shown.}
    \label{fig:limit_mig}
\end{figure*}

The signal generated in each of the inelastic scattering processes described in Sections~\ref{ssec:brems} and \ref{ssec:migdal} is a differential rate. Therefore, and in contrast to the absorption processes, a different energy dependence of $\sigma_{\mathrm{p.e.}}$ alters the shape of the expected spectra, not just its scale. This can be seen in Fig.~\ref{fig:sig_comp} comparing the differential rates based on the fitted and nominal $\sigma_{\mathrm{p.e.}}$ curves and before detector effects are applied. The corresponding quantized dark matter signal spectra expected for the Migdal process in the HVeV detector are shown in Fig.~\ref{fig:R2spec}. The ionization model used for the quantization follows Ref.~\cite{Amaral:2020ryn} with a Fano factor of 0.155. As can be seen from both figures, the difference is largest at low energies and negligible at higher energies. At the same time the data spectrum observed in single-electron sensitive devices is typically highest at lowest energies (as e.g. in Fig.~\ref{fig:R2spec}) which is largely due to a non-zero dark current \cite{Amaral:2020ryn}. As a result these types of experiments are most sensitive to Bremsstrahlung and Migdal processes at energies that are not affected by the presented difference in the $\sigma_{\mathrm{p.e.}}$ curves and thus the respective exclusion limits on $\sigma^\mathrm{SI}_n$ will be minimally affected, if at all. The example dark matter signal spectra shown in Fig.~\ref{fig:R2spec} are both scaled to a cross section of $\sigma_n^\mathrm{SI}=1.36\times 10^{-30} $\,cm$^2$, where the strongest limit is set for both curves on the forth electron-hole pair peak. The difference in the first electron-hole pair peak, as shown in the figure inset, does not have an impact on the limit.

For different dark masses different electron-hole pair peaks can provide the highest sensitivity to a potential dark matter signal. However, for none of the dark matter masses probed with the HVeV data does the first electron-hole pair peak provide the highest sensitivity. Figures~\ref{fig:limit_mig} compare the 90\% C.L. limits computed using either the nominal or fitted $\sigma_{\mathrm{p.e.}}$ curve. The limits that are set based on HVeV data are virtually identical. To decouple the discussion from a specific data set and to understand the maximum impact the fitted $\sigma_{\mathrm{p.e.}}$ curve can have on the limits, the calculations were repeated with an idealized zero-background toy experiment as described in Sec.~\ref{ssec:data}. Using only the first electron-hole pair peak in this calculation (see Sec.~\ref{ssec:poisson}), a limit weaker by about 10\% (5\%) for the Migdal (Bremsstrahlung) signal model is observed over the entire dark matter mass range when the fitted $\sigma_{\mathrm{p.e.}}$ curve is used instead of the nominal one. Whether or not this effect is relevant depends on the size of the various uncertainties that may exist in a respective measurement. Typical experimental uncertainties include, but are not limited to, uncertainties on the energy resolution, on the charge trapping and impact ionization fraction, and on the signal efficiency. A systematic uncertainty that is of particular interest for single electron-hole pair sensitive devices is introduced with the ionization model used to quantize the expected dark matter signal spectrum (as shown in Fig.~\ref{fig:R2spec}) and the respective Fano factor that is applied \cite{Agnese_2018, Amaral:2020ryn}. Which uncertainty, including the one introduced by the $\sigma_{\mathrm{p.e.}}$ differences, is the dominating one is highly dependent on the experiment.

It should be noted that the presented limits on $\sigma^\mathrm{SI}_n$ do not take into account an upper limit that would be caused by dampening from the atmosphere (the atmosphere plus the Earth) for dark matter coming from the upper (lower) hemisphere in case of a surface-operated experiment. This means that these results could be in a region that is already excluded when accounting for these dampening effects. 
Also the lower limits themselves shown in Fig.~\ref{fig:limit_mig} are not corrected for atmosphere and Earth shielding effects \cite{Kavanagh_2017}. This means in particular that the limit results in Fig.~\ref{fig:limit_mig} should not be taken as actual constraints on $\sigma_n^\mathrm{SI}$ derived from the HVeV Run 2 data in Ref.~\cite{Amaral:2020ryn}. They are only intended to demonstrate the potential effect of $\sigma_{\mathrm{p.e.}}$ on such limits. The dampening effects were not taken into account because they are beyond the purpose of this paper. The conclusions from the presented results using the HVeV Run 2 data and the zero background toy experiment are unaffected.


\section{\label{sec:conclusion}Conclusion}

For both dark photon absorption and ALP absorption, the use of the fitted $\sigma_{\mathrm{p.e.}}$ curve that accounts for the temperature dependence over the nominal $\sigma_{\mathrm{p.e.}}$ curve, as determined in Ref.~\cite{pexsec_stanford}, results in an exclusion limit up to two times greater at masses below 4\,eV/$c^2$. The difference becomes asymptotically greater close to the Si band gap. For the dark photon exclusion limit, the fitted $\sigma_{\mathrm{p.e.}}$ curve produces a slightly stronger limit above about 3.5\,eV/$c^2$ after applying an in-medium correction. These findings are based on SuperCDMS HVeV Run\,2 data but are expected to be qualitatively applicable to comparable experiments like SENSEI and DAMIC \cite{SENSEI2019, DAMIC2019}.

In case of the Bremsstrahlung and Migdal interaction channels, using either the nominal or fitted $\sigma_{\mathrm{p.e.}}$ curve in the dark matter signal model produces no significant difference in the exclusion limits set on SuperCDMS HVeV Run\,2 data. This result, however, is dependent on whether or not the first electron-hole pair peak in the respective data spectrum contributes to the limit result. Calculating the limits on an idealized experiment with no events and perfect efficiency, as well as only setting the limit on the first electron-hole pair peak, maximizes the difference that could be observed in the limits when using the different $\sigma_{\mathrm{p.e.}}$ curves. For the  Bremsstrahlung and Migdal signal models, the difference in the exclusion limits in this scenario are about a factor of 1.05 and 1.1, respectively.

In general, this analysis highlights the importance of considering the temperature effects of the photoelectric absorption cross section when conducting analyses on low-mass dark matter candidates that depend on this parameter. For dark matter absorption, the effect is clear and significant at low masses. For inelastic DM-nucleus scattering, the effect is more subtle and potentially negligible. Only in cases in which the experimental sensitivity to these interaction channels is driven by the first electron-hole pair peak can this effect become noticeable.

\begin{acknowledgments}
We would like to thank Rouven~Essig and Josef~Pradler for helpful discussions about the use of the photoelectric absorption cross section parameter in the Migdal and Bremsstrahlung interaction models. We would further like to thank Matthew~Dolan, Timon~Emken, Masahiro~Ibe, Felix~Kahlhoefer, Chris~McCabe, Wakutaka~Nakano, Jayden~Newstead, Yutaro~Shoji, and Kazumine~Suzuki for their useful discussions on the Migdal Effect, its normalization and its application in silicon. We also thank Betty~Young and Steven~Yellin for useful discussions and feedback on the manuscript. This work was supported by the Deutsche Forschungsgemeinschaft (DFG) under Project No. 420484612 and Germany’s Excellence Strategy - EXC 2121 ``Quantum Universe" – 390833306, by the U.S. Department of Energy and by the National Science Foundation (Grant No. 1707704), and by the Canada First Research Excellence Fund through the Arthur~B.~McDonald Canadian Astroparticle Physics Research Institute. This document was prepared by using resources of the Fermi National Accelerator Laboratory (Fermilab), a U.S. Department of Energy, Office of Science, HEP User Facility. Fermilab is managed by Fermi Research Alliance, LLC (FRA), acting under Contract No. DE-AC02-07CH11359. SLAC is operated under Contract No. DEAC02-76SF00515 with the U.S. Department of Energy.
\end{acknowledgments}



\bibliography{refs}

\end{document}